\documentclass[10pt,a4paper,twoside]{article}
\usepackage{epsfig}
\usepackage{baltlat6}
\usepackage{array}
\usepackage{here}
\usepackage{natbib}
\begin{document}

\vspace{0.5mm}
\setcounter{page}{1}
\vspace{8mm}

\def\aj{AJ}%
\def\actaa{Acta Astron.}%
\def\araa{ARA\&A}%
\def\apj{ApJ}%
\def\apjl{ApJ}%
\def\apjs{ApJS}%
\def\ao{Appl.~Opt.}%
\def\apss{Ap\&SS}%
\def\aap{A\&A}%
\def\aapr{A\&A~Rev.}%
\def\aaps{A\&AS}%
\def\azh{AZh}%
\def\mnras{MNRAS}%
\def\memras{MmRAS}%
\def\na{New A}%
\def\pra{Phys.~Rev.~A}%
\def\prb{Phys.~Rev.~B}%
\def\prc{Phys.~Rev.~C}%
\def\prd{Phys.~Rev.~D}%
\def\pre{Phys.~Rev.~E}%
\def\prl{Phys.~Rev.~Lett.}%
\def\pasp{PASP}%
\def\nat{Nature}%
\def\bain{Bull.~Astron.~Inst.~Netherlands}%
\def\Mpc{$h^{-1}$~{\rm  Mpc}}

\titlehead{Baltic Astronomy, vol.\,20,   , 2011}

\titleb{DARK MATTER}

\begin{authorl}
\authorb{J. Einasto}{1}
\end{authorl}

\begin{addressl}
\addressb{1}{Tartu Observatory, 61602 T\~oravere, Estonia; einasto@aai.ee}

\end{addressl}

\submitb{Received: 2011 July 6; accepted: 2011 July 3}

\begin{summary} 
  I give a review of the development of the concept of dark matter.
  The dark matter story passed through several stages from a minor
  observational puzzle to a major challenge for theory of elementary
  particles.  Modern data suggest that dark matter is the dominant
  matter component in the Universe, and that it consists of some
  unknown non-baryonic particles.  Properties of dark matter particles
  determine the structure of the cosmic web.
\end{summary}

\begin{keywords} 
Dark matter, galaxies, clusters of galaxies, large-scale
 structure of the Universe
\end{keywords}

\resthead{Dark matter}
{J. Einasto}

\sectionb{1}{INTRODUCTION}

The possible presence of invisible but gravitating dark matter has
been studied already almost hundred years. First attempts to derive
the total density of matter in the Solar vicinity were made by
\citet{Opik:1915}, \citet{Kapteyn:1922}, \citet{Jeans:1922}, and
\citet{Oort:1932}.  Results were different: \citet{Opik:1915},
\citet{Kapteyn:1922} and \citet{Oort:1932} found that the total
density of matter can be explained by known stellar populations, if a
reasonable extrapolation of faint dwarf stars is taken into
account. In contrast, \citet{Jeans:1922} found that there must be two
dark stars to each bright star. This discussion continued until the
end of 20th century.

A much larger discrepancy between the masses of visible objects and
the total masses of stellar systems they belong to, was discovered by
\citet{Zwicky:1933}.  He concluded that, in order to hold galaxies
together in the cluster, the cluster must contain huge amounts of dark
(invisible) matter.

Initially  no distinction between local dark matter in the solar
vicinity and global dark matter in clusters of galaxies was
made. The realisation, that these two types of dark matter have very
different properties and nature came from the detailed study of
galactic models \citep{Einasto:1974a}.

In this talk I have used my recent review of the dark matter \citep{Einasto:2009zr}.

\sectionb{2}{LOCAL DARK MATTER}

The dynamical density of matter in the solar vicinity can be estimated
using vertical oscillations of stars around the galactic plane.
\citet{Opik:1915} found that the summed contribution of all known
stellar populations and interstellar gas is sufficient to explain the
vertical oscillations of stars -- in other words, there is no need to
assume the existence of a dark population.  Similar analyses were made
by \citet{Kapteyn:1922} and \citet{Jeans:1922}, who used the term
``Dark Matter'' to denote the invisible matter.  Kapteyn found for the
dynamical density of matter near the Sun 0.099~M$_\odot$/pc$^3$, Jeans
got 0.143 in the same units.

\citet{Oort:1932} analysis suggested that the total density is about
0.092~M$_\odot$/pc$^3$, and the density of stars, including expected
number of white dwarfs, is approximately equal to the dynamical
density. He concluded that the total mass of nebulous or meteoric dark
matter near the Sun is very small.

\citet{Kuzmin:1952, Kuzmin:1955} and his students \citet{Eelsalu:1959}
and \citet{Joeveer:1972, Joeveer:1974} confirmed the earlier results
by \"Opik, Kapteyn and Oort. A number of other astronomers, including
more recently \citet{Oort:1960} and \citet{Bahcall:1984, Bahcall:1987},
found results in agreement with the Jeans result.  Their results mean
that the amount of invisible matter in the solar vicinity should be
approximately equal to a half of the amount of visible matter.

Modern data by \citet{Kuijken:1989b, Gilmore:1989} have confirmed the
results by \citet{Kapteyn:1922}, \citet{Oort:1932},
\citet{Kuzmin:1952, Kuzmin:1955} and his collaborators.  Thus we come
to the conclusion that {\em there is no evidence for the presence of
  large amounts of dark matter in the disk of the Galaxy}.  If there
is some invisible matter near the galactic plane, then its amount is
small, of the order of 15 percent of the total mass density. 

\sectionb{3}{GLOBAL DARK MATTER}

\citet{Zwicky:1933,Zwicky:1937uz} measured radial velocities of
galaxies in the Coma cluster of galaxies, and calculated the mean
random velocities in respect to the mean velocity of the
cluster. Galaxies move in clusters along their orbits; the orbital
velocities are balanced by the total gravity of the cluster.  Zwicky
found that orbital velocities are almost a factor of ten larger than
expected from the summed mass of all galaxies belonging to the
cluster.

A certain discrepancy was detected between masses of individual
galaxies and masses of pairs and groups of galaxies
\citep{Holmberg:1937uq, Page:1952, Page:1959, Page:1960}.
These determinations yield for the mass-to-light ratio (in blue light)
the values $M/L_B = 1 \dots 20$ for spiral galaxy dominated pairs, and
$M/L_B = 5 \dots 90$ for elliptical galaxy dominated pairs. These
ratios are larger than found from local mass indicators of galaxies
(velocity dispersions at the centre and rotation curves of spiral
galaxies).

\citet{Kahn:1959} paid attention to the fact that most galaxies have
positive redshifts, only the Andromeda galaxy (M31) has a negative
redshift of about 120 km/s, directed toward our Galaxy. This fact can
be explained, if both galaxies form a physical system. From the
approaching velocity, the mutual distance, and the time since passing
the perigalacticon (taken equal to the present age of the Universe),
the authors calculated the total mass of the double system. They found
that $M_{tot} \geq 1.8 \times 10^{12}$  M$_{\odot}$.  The conventional
masses of the Galaxy and M31 were estimated to be of the order of $2
\times 10^{11}$ M$_{\odot}$.  In other words, the authors found
evidence for the presence of additional mass in the Local Group of
galaxies.

Information of masses of individual galaxies come from their rotation
velocities. \citet{Roberts:1966dt} made a 21-cm hydrogen line survey
of M31 using the National Radio Astronomy Observatory large 300-foot
telescope. He found that the rotation velocity curve at large radii is
flat, i.e., velocity is almost constant. From the comparison of the
light distribution with the rotation curve the local value of the
mass-to-luminosity can be calculated. He found in the outer regions a
mass-to-light ratio $\sim 250$.  A similar high value was found for
the edge-on S0 galaxy NGC 3115 by
\citet{Oort:1940}. \citet{Rubin:1970} and \citet{Roberts:1973fb}
derived the rotation curve of M31 up to a distance $\sim 30$ kpc,
using optical and radio data, respectively.  The rotation speed rises
slowly with increasing distance from the centre of the galaxy and
remains almost constant over radial distances of 16--30 kpc.

Two possibilities were suggested to explain flat rotation curves of
galaxies.  One possibility is to identify the observed rotation
velocity with the circular velocity. In this case an explanation for a
very high local $M/L$ should be found.  To explain this phenomenon it
was suggested that in outer regions of galaxies low-mass dwarf stars
dominate \citep{Oort:1940, Roberts:1975}.  The other possibility is to
assume that in the periphery of galaxies there exist non-circular
motions which distort the rotation velocity.

\sectionb{4}{GALACTIC MODELS}

Classical models of elliptical galaxies were found from luminosity
profiles and were calibrated using either central velocity dispersions, or
motions of companion galaxies.  Models of spiral galaxies were
constructed using rotation velocities.  A natural generalisation of
classical galactic models is the use of all available observational
data -- photometric data on the distribution of colour and light, and
kinematical data on the rotation and/or velocity dispersion.  Further,
it is natural to include into models data of all major stellar
populations, such as the bulge, the disk, the halo, as well as the
flat population in spiral galaxies, consisting of young stars and
interstellar gas.

All principal descriptive functions of galaxies (circular velocity,
gravitational potential, projected density) are simple integrals of
the spatial density. Therefore it is natural to apply for the spatial
density $\rho(a)$ of galactic populations a simple generalised
exponential expression \citep{Einasto:1965}:
\begin{equation}
\rho(a) = \rho(0) \exp\left(-(a/a_{0})^{1/N}\right),
\label{explaw}
\end{equation}
where $a$ is the semi-major axis of the isodensity ellipsoid, $a_0$ is
the effective radius of the population, and $N$ is a structural
parameter, determining the shape of the density profile. This
expression (called the Einasto profile) can be used for all galactic
populations, including dark halos. The case $N=4$ corresponds to the
de Vaucouleurs density law for spheroidal populations, $N=1$
corresponds to the exponential density law for disk.  

To combine photometric and kinematic data, mass-to-light ratios of
galactic populations are needed. Luminosities and colours of galaxies
in various photometric systems result from the physical evolution of
stellar populations that can be modelled.  Detailed models of the
physical and chemical evolution of galaxies were constructed by
\citet{Tinsley:1968}.  Combined population and physical evolution
models were calculated for a representative sample of galaxies by
\citet{Einasto:1972}. It is natural to expect, that in similar
physical conditions the mass-to-luminosity ratio $M_i/L_i$ of the
population $i$ has similar values in different stellar systems (star
clusters, galactic populations). Thus star clusters and central cores
of galaxies can be used to estimate $M_i/L_i$ values for the main
galactic populations.

\sectionb{5}{GALACTIC CORONAS}

Results of these calculations were reported at the First European
Astronomy Meeting by \citet{Einasto:1974a}.  The main conclusion was:
it is impossible to reproduce the rotation data by known stellar
populations only.  The only way to eliminate the conflict between
photometric and rotational data was {\em to assume the presence of an
  unknown almost spherical population with a very high value of the
  mass-to-light ratio, large radius and mass}.  To avoid confusion
with the conventional stellar halo, the term ``corona'' was suggested
for the massive population.  Thus, the detailed modelling confirmed
earlier results obtained by simpler models.  But here we have one
serious difficulty -- no known stellar population has so large a $M/L$
value.

Additional arguments for the presence of a spherical massive
population in spiral galaxies came from the stability criteria against
bar formation, suggested by \citet{Ostriker:1973}. Their numerical
calculations demonstrated that flat systems become rapidly thicker and
evolve to a bar-like body. In real spiral galaxies a thin population
exists, and it has no bar-like form.  To remain stable galaxies must
have a massive spherical halo. 

The rotation data available in the early 1970s allowed the determination
the mass distribution in galaxies up to their visible edges.  In order
to find how large and massive galactic coronas or halos are, more
distant test particles are needed.  If halos are large enough, then in
pairs of galaxies the companion galaxies are located inside the halo,
and their relative velocities can be used instead of the galaxy
rotation velocities to find the distribution of mass around giant
galaxies.   This test was made by \citet{Einasto:1974}. A similar
study was made independently by \citet{Ostriker:1974}. Our results
were first discussed in the Caucasus Winter School on Cosmology in January
1974 and in the Tallinn Conference on Dark Matter in January 1975
\citep{Doroshkevich:1975}. 

The mass of galactic coronas exceeds the mass of populations of known
stars by one order of magnitude. According to new estimates the total
mass density of matter in galaxies is 20\% of the critical
cosmological density. The data suggest that all giant galaxies have
massive halos/coronas, thus dark matter must be the dynamically
dominating population in the whole Universe.

Initially the presence of massive coronas/halos was met with
scepticism.  In  the Third European Astronomical Meeting 
the principal discussion was between the supporters of the classical
paradigm with conventional mass estimates of galaxies, and of the new
one with dark matter.  The major arguments supporting the classical
paradigm were summarised by \citet{Materne:1976}. Their most serious
argument was:

 {\em ``Big Bang nucleosynthesis suggests a low-density
  Universe with the density parameter $\Omega \approx 0.05$; the
  smoothness of the Hubble flow also favours a low-density Universe.''}

Additional observational data gave strong support to the presence of
massive coronas/halos. Available rotation data were summarised by
\citet{Roberts:1975}. Extended rotation curves were available for 14
galaxies. In all galaxies the local mass-to-light ratio in the
periphery reached values over 100 in solar units.  \citet{Rubin:1978,
  Rubin:1980} measured optically the rotation curves of galaxies at
very large galactocentric distances.  \citep{Bosma:1978} measured
rotation data for 25 spiral galaxies with the Westerbork Synthesis
Radio Telescope.  Both results suggested that practically all spiral
galaxies have extended flat rotation curves.
 
Another very important measurement was made by \cite{Faber:1976,
  Faber:1977, Faber:1979}.  They measured the central velocity
dispersions for 25 elliptical galaxies and the rotation velocity of
the Sombrero galaxy, just outside the main body of the bulge.  Their
data yielded for the bulge of the Sombrero galaxy a mass-to-light
ratio $M/L=3$, and for the mean mass-to-light ratios for elliptical
galaxies about 7. These results showed that the mass-to-light ratios
of stellar populations in spiral and elliptical galaxies are similar
for a given colour, and the ratios are much lower than accepted in
earlier studies based on the dynamics of groups and clusters.  In
other words, high mass-to-light ratios of groups and clusters of
galaxies cannot be explained by visible galactic populations.

The distribution of the mass in clusters can be determined if the
density and the temperature of the intra-cluster gas are known. These
data can be measured by the Einstein X-ray orbiting observatory. The
mass of Coma, Perseus and Virgo clusters was calculated from X-ray
data by \citet{Bahcall:1977, Mathews:1978a}. The results confirmed
previous estimates of masses made with the virial method using
galaxies as test particles.

Finally, masses of clusters of galaxies can be measured using
gravitational lensing of distant galaxies by clusters. The masses of clusters
of galaxies determined using this method, confirm the results obtained by the
virial theorem and the X-ray data \citep{Fischer:1997, Fischer:1997a}.

Earlier suggestions on the presence of mass discrepancy in galaxies
and galaxy systems had been ignored by the astronomical community.
This time new results were taken seriously. However, it was still not
clear how to explain the controversy of the Big Bang nucleosynthesis
and the smoothness of the Hubble flow, discussed by \citet{Materne:1976}.

\sectionb{6}{THE NATURE OF DARK MATTER}

The local dark matter is baryonic (low-mass stars or Jupiter-like
objects), since non-baryonic matter is dissipationless and cannot form
a highly flattened population.

The nature of the global dark matter has been a subject of discussion
for long time.  Initially it was suggested that in outer regions of
galaxies low-mass dwarf stars dominate \citep{Oort:1940,
  Ostriker:1974, Roberts:1975}.  However, the stellar nature of
galactic coronas/halos meets several difficulties.  Coronas have 
larger dimensions than all known stellar populations, thus from
hydrostatic equilibrium condition coronal stars must have much higher
velocity dispersions than other populations.  No fast-moving stars
were found \citep{Jaaniste:1975}. If the hypothetical population is of
stellar origin, it must be formed much earlier than all known
populations, because known stellar populations of different age and
metallicity form a continuous sequence of kinematical and physical
properties, and there is no place where to include this new population
\citep{Einasto:1974a}.  It is known that star formation is not an
efficient process -- usually in a contracting gas cloud only about
1~\%~ of the mass is converted to stars. Thus we have a problem how to
convert, in an early stage of the evolution of the Universe, a large
fraction of the primordial gas into this population of dark stars.

The nature of dark matter was the basic problem discussed in the
Caucasus Winter School and in the Dark Matter Conference in Tallinn
\citep{Doroshkevich:1975}.  \citet{Silk:1974,Komberg:1975a} showed
that gaseous coronas of galaxies and clusters cannot consist of
neutral gas since the intergalactic hot gas would ionise the coronal
gas.  A corona consisting of hot ionised gas would be observable.  A
fraction of the coronal matter around galaxies and in groups and
clusters of galaxies consists indeed of the X-ray emitting hot gas,
but the amount of this gas is not sufficient to explain the flat
rotation curves of galaxies \citep{Turner:2003}.

The baryonic nature (stars, gas) of the dark matter contradicts also
the nucleosynthesis constraints mentioned already by
\citet{Materne:1976}.  A third very important observation was made
which caused doubts to the baryonic matter as the dark matter
candidate. The Cosmic Microwave Background radiation temperature and
density fluctuations are much lower than the theoretically predicted
limit $10^{-3}$ (see, for instance \citet{Parijskij:1978}).

Then astronomers considered the possible existence of non-baryonic
particles, such as heavy neutrinos. This suggestion was made
independently by several astronomers (\citet{Szalay:1976} and others).
If dark matter consists of heavy neutrinos, then this helps to explain
the paradox of small temperature fluctuations of the cosmic microwave
background radiation.  Dark matter starts to condense at early epoch and forms
potential wells, after the recombination  baryonic matter flows into these wells and forms
galaxies. However, numerical simulations of the formation
of the structure of the neutrino-dominated dark matter Universe
demonstrated, that in this case only supercluster-scale systems can
form (see below).

\sectionb{7}{DARK MATTER AND LARGE-SCALE STRUCTURE OF THE UNIVERSE}

After my talk at the Caucasus Winter School Zeldovich offered me
collaboration in the study of the universe.  He was developing a
theory of formation of galaxies (the pancake theory, \citet{Zeldovich:1970}).
 A  hierarchical clustering theory was suggested by \citet{Peebles:1971}.  Zeldovich
asked for our help in solving the question: can we find some
observational evidence which can be used to discriminate between these
theories?

In solving the problem we used our previous experience in the study of
galactic populations: kinematical and structural properties of
populations hold the memory of their previous evolution. Random velocities of
galaxies are of the order of several hundred km/s, thus during the
whole lifetime of the Universe galaxies have moved from their place of
origin only by about 1~\Mpc\ (the Hubble constant is used 
in  units of $H_0 = 100~h$ km~s$^{-1}$~Mpc$^{-1}$).  In other words
-- if there exist some regularities in the distribution of galaxies,
these regularities must reflect the conditions in the Universe during
the formation of galaxies.  

Thus we had a leading idea how to solve the problem of galaxy
formation: {\em We have to study the distribution of galaxies on
  larger scales}. The three-dimensional distribution of galaxies,
groups and clusters of galaxies can be visualised using
wedge-diagrams, invented just when we started our study. We prepared
relatively thin wedge diagrams, and plotted in the same diagram
galaxies, as well as groups and clusters of galaxies.  In these
diagrams regularity was clearly seen: {\em isolated galaxies and
  galaxy systems populated identical regions, and the space between
  these regions was empty}. This picture was quite similar to the
distribution of test particles in a numerical simulation of the
evolution of the structure of the Universe.

We reported our results \citep{Joeveer:1978a}  at the IAU
symposium on Large-Scale Structure of the Universe in Tallinn 1977,
the first conference on this topic. The main results were: 

\begin{enumerate}

\item galaxies, groups and clusters of galaxies are not randomly
distributed but form chains, converging in superclusters;

\item the space between galaxy chains contains almost no galaxies and
forms holes (voids) of diameter up to $\approx 70$~\Mpc;

\item the whole picture of the distribution of galaxies and clusters
resembles cells of a honeycomb, rather close to the picture predicted
by \citet{Zeldovich:1978}.
\end{enumerate}

However, some important differences between the Zeldovich pancake
model and observations were detected.  First of all, there exists a
rarefied population of test particles in voids absent in real
data. This was the first indication for the presence of biasing in
galaxy formation -- there is primordial gas and dark matter in voids,
but due to low-density no galaxy formation takes place here
\citep{Joeveer:1978b, Einasto:1980}. The second difference lies in the
structure of galaxy systems in high-density regions: in the model
large-scale structures (superclusters) have rather diffuse forms, real
superclusters consist of multiple intertwined filaments
\citep{Zeldovich:1982, Oort:1983, Bond:1996}.

The difficulties of the neutrino-dominated model became evident in
early 1980s.  A new scenario was suggested by \citet{Blumenthal:1982}
and others, where hypothetical particles like axions, gravitinos or
photinos play the role of dark matter.  Numerical simulations of
structure evolution for neutrino and axion-gravitino-photino-dominated
universe were made and analysed by \citet{Melott:1983}.  All
quantitative characteristics (the connectivity of the structure, the
multiplicity of galaxy systems, the correlation function) of this new
model fit the observational data well.  This model was called the Cold
Dark Matter (CDM) model, in contrast to the neutrino-based Hot Dark
Matter model.  Presently the CDM model with some modifications is the
most accepted model of the structure evolution. The properties of the
Cold Dark Matter model were analysed in detail by
\citet{Blumenthal:1984}.

The modern cosmological paradigm includes Dark Energy as the basic
component of the matter/energy content of the Universe. Direct
observational evidence for the presence of Dark Energy comes from
distant supernova observations \citep{Perlmutter:1999,Riess:1998} and
CMB observations. The Wilkinson Microwave Anisotropy Probe (WMAP)
satellite allowed the measurement of the CMB radiation and its power
spectrum with a much higher precision \citep{Spergel:2003}.  The
position of the first maximum of the power spectrum depends on the
total matter/energy density. Observations confirm the theoretically
favoured value 1 in units of the critical cosmological density.
Combined CMB, supernova and large-scale distribution data yield for
the density of baryonic matter, $\Omega_b = 0.041$, the dark
matter density $\Omega_{DM} = \Omega_m - \Omega_b = 0.23$, and the
dark energy density $\Omega_\Lambda = 0.73$.  These parameters imply
that the age of the Universe is $13.7 \pm 0.2$ gigayears.

Dark energy act as a repulsive force, thus the Universe is presently
expanding with an increasing speed.  Dark energy also has the effect
of freezing the cosmic web.  This explains the smoothness of the
Hubble flow. The nature of dark matter particles and dark energy is 
still unknown.

\sectionb{8}{CONCLUSIONS}	

\begin{itemize}
\item The discovery of Dark Matter was the result of combined study of
  galaxies, clusters and their distribution.
\item Dark Matter Story is a typical scientific revolution
  \citep{Tremaine:1987}.
\item Evidence for dark matter has been collected independently in
  many centres.
\item There are 2 dark matter problems:  dark matter in the Galaxy disk,
  and dark matter around galaxies and clusters.
\item  Dark matter in the disk is baryonic (faint stars or
  Jupiters). The amount is small.
\item  Dark matter around galaxies is non-baryonic Cold Dark
  Matter. It constitutes about 0.25 of critical cosmological density.
\item Non-baryonic dark matter is needed to start early enough gravitational
  clustering to form structure. This solves the Big-Bang
  Nucleosynthesis controversy.
\end{itemize}


\end{document}